\documentclass[12pt]{article}
\usepackage{bezier}
\usepackage{amssymb}
\usepackage{epsfig}

\newcommand{\nc}{\newcommand}
\nc{\be}{\begin{equation}}
\nc{\ee}{\end{equation}}

\nc{\mlgraph}{{multiple-line graph }}
\nc{\mlgraphs}{{mul\-tiple\--line graphs }}
\nc{\Mlgraph}{{Multiple-line graph }}

\nc{\mldiagram}{{multiple-line diagram }}
\nc{\mldiagrams}{{multiple-line diagrams }}

\nc{\mlmoment}{{multiple-line moment}}
\nc{\mlmoments}{{multiple-line moments}}

\nc{\bea}{\begin{eqnarray}}
\nc{\eea}{\end{eqnarray}}
\nc{\bela}{\begin{eqnarray*}}
\nc{\eela}{\end{eqnarray*}}

\nc{\eqn}[1]{{(\ref{#1})}}
\nc{\cA}{{\cal A}}
\nc{\cB}{{\cal B}}
\nc{\cC}{{\cal C}}
\nc{\cD}{{\cal D}}
\nc{\cE}{{\cal E}}
\nc{\cF}{{\cal F}}
\nc{\cG}{{\cal G}}
\nc{\cH}{{\cal H}}
\nc{\cI}{{\cal I}}
\nc{\cJ}{{\cal J}}
\nc{\cK}{{\cal K}}
\nc{\cL}{{\cal L}}
\nc{\cM}{{\cal M}}
\nc{\cN}{{\cal N}}
\nc{\cO}{{\cal O}}
\nc{\cP}{{\cal P}}
\nc{\cQ}{{\cal Q}}
\nc{\cR}{{\cal R}}
\nc{\cS}{{\cal S}}
\nc{\cT}{{\cal T}}
\nc{\cU}{{\cal U}}
\nc{\cV}{{\cal V}}
\nc{\cW}{{\cal W}}
\nc{\cX}{{\cal X}}
\nc{\cY}{{\cal Y}}
\nc{\cZ}{{\cal Z}}

\nc{\simo}[1]{{\stackrel{#1}{\simeq}}}
\nc{\geqo}[1]{{\stackrel{#1}{\geq}}}
\nc{\geo}[1]{{\stackrel{#1}{>}}}
\nc{\guo}[1]{{\stackrel{#1}{\succ}}}

\nc{\rbo}{\raisebox}
\nc{\RR} {\rangle \! \rangle}
\nc{\LL} {\langle \! \langle}
\nc{\rmi}[1]{{\mbox{\small #1}}}
\nc{\eq}{eq.~}
\nc{\nr}[1]{(\ref{#1})}
\nc{\ul}{\underline}
\nc{\mc}{\multicolumn}
\nc{\todo}[1]{\par\noindent{\bf $\rightarrow$ #1}}

\nc{\cu}{{\cal u}}

\title{
\begin{flushright} {\small $\begin{array}{ l } \mbox{IUB--TH-038} \\
    \mbox{} \end{array} $}
 \end{flushright}
\vskip10pt
Functional Complexity Measure for Networks
}
\author{Hildegard~Meyer-Ortmanns\\
School of Engineering and Science\\
International University Bremen\\
P.O.Box 750561\\
D-28725 Bremen, Germany \\
e-mail: h.ortmanns@iu-bremen.de}

\begin{document}

\maketitle

\begin{abstract}
\setlength{\baselineskip}{1pt}
\noindent
We propose a complexity measure which addresses the functional flexibility of 
networks. It is conjectured that the functional flexibility is reflected in the
topological ``diversity'' of the assigned graphs, resulting from a 
resolution of their vertices and a rewiring of their edges under 
certain constraints. The application will be a classification of 
networks in artificial or biological systems, where functionality plays 
a central role.
\\
\noindent{\bf Keywords}: networks, complexity measure, functional complexity
\end{abstract}

\section{Introduction}
Complexity measures have been proposed as measures for computational, 
statistical, or structural complex features in various contexts, 
for a review see \cite{wackerbauer}. 
A complexity measure for patterns, for example, arising in chaotic
systems has been proposed in \cite{grassberger}. It is a measure theoretic concept and applies to ensembles of
patterns. It is natural in the sense that it reflects the intuitive notion of a complex pattern being neither
completely random nor completely regular, but having some structure instead. Complexity of hierarchical systems 
has been studied in \cite{huberman}. The complexity measure there has the property of isolating the most diverse 
trees as the ones with maximal complexity. Intuitively one would expect that the complexity of a hierarchy is related
to its diversification, i.e. to the number of non-isomorphic subtrees found at that level. The proposals of
\cite{huberman} reproduce this expectation.

Our measure of complexity refers to more general systems described by networks with underlying graphs that have
not necessarily a tree-like structure. Also we do not address the variety 
and the complexity of patterns which may be 
produced from nets with simple topologies by adjusting the dynamical couplings attached to the edges.
We are interested in the functional complexity of networks, natural and 
artificial ones, whose design is often optimized in view of a variety of possible functions. In particular the
concept of functionality applies to networks in life science and in information science. Network motifs have been
studied as characteristic building blocks for complex networks \cite{motifs}. As a result, motifs shared by ecological
food webs are specifically different from genetic networks. More generally, 
it has been found in \cite{motifs} that motifs
in networks of information processing are typically distinct from networks of
energy transporting. Information processing
may refer to nets as diverse as those of gene regulation, neurons and electronic circuits, for example. The 
overall conclusion is that frequently repeated motifs should 
represent certain functions. 

Our proposal of a complexity measure
is in a similar spirit. We believe that the functional diversity of a network is determined by the topological
"diversity" of the corresponding graph which results from rewiring the lines 
(we use ``lines'' rather than ``edges'') or breaking up the vertices (nodes) 
into pieces in all allowed ways. To make these concepts well defined, we use
the framework of DLCE-graphs which have been proposed in \cite{hmoreisz} in
a very different context. The former context was a bookkeeping of 
all analytic contributions to a given order in a generalized high-temperature
expansion. Differently from the graphs that are mostly considered in connection with dynamical networks, DLCE-graphs 
have two types of vertices along with two types of connectivities, and, in general, more than one line connecting  
two vertices. In a superimposed dynamics, dynamical variables are associated with both, the vertices and the lines.
We make use of a special operation on these graphs, later called the resolution of vertices.

In section 2 we define DLCE-graphs including some notions of a multiple-line graph theory as it was introduced in 
\cite{hmoreisz}. In section 3 we propose the measure of functional complexity 
and illustrate its use with some
examples. Section 4 adds a summary of the physical background where DLCE-graphs were studied first. This section
may be skipped by the reader who is not interested in the expansions where these graphs naturally occur. Section 5
gives the summary and a short outlook.


\section{DLCE-graphs and Networks}

Motivated by spin glass dynamics on lattices (with regular, random or other network topologies) we introduce so-called 
DLCE-graphs. DLCE stands for Dynamical Linked Cluster Expansions. These are systematic high temperature expansions 
generalized to systems where the spins and their couplings both have their own dynamics and interact with each other.
For further background on DLCE we refer to section 4 and \cite{hmoreisz}. Since the underlying dynamics of DLCE and
the corresponding graphical expansion are quite generic, we generalize the usual definition of graphs to DLCE-graphs also in this context.

DLCE-graphs have two types of connectivity. Vertices are connected via lines 
and lines may be additionally
connected via a different kind of vertex, here called beam. Therefore we have two kind of vertices. When we speak of vertices in the following, we mean the
first type, and use ``beams'' for the second type. Two vertices may
be connected via more than one line, but two lines may be connected at most 
via one beam, because each line belongs
to exactly one beam. (In spin glasses the spins would 
live on vertices, the spin couplings on lines (there usually called
links or bonds), and the self-interaction of the couplings 
induces the beams.) If a beam connects $m$
lines, we call the resulting structure an m-line. A one-line coincides with a "usual" line. A multiple-line graph 
theory has been developed in \cite{hmoreisz}. Here we introduce only notions 
that we later need for our applications.
We distinguish between internal and external lines: internal lines get vertices attached to both of their ends.
For external lines we distinguish two types, a so-called $\Phi$-line with a vertex attached to only one endpoint, and a
$U$-line with no vertices attached to its endpoints. (In connection with spin glasses the external $\Phi$-and $U$-lines
correspond to n-point correlation functions of $\Phi$ (spin) or $U$ (coupling)-fields.) In a generic network the external
lines correspond to input-or output channels, where both may have $\Phi$-or $U$-variables attached, respectively.

Let us now define in detail the notion of a DLCE-graph and the topological equivalence of two such graphs. 
A DLCE-graph is a strucure
\be
\Gamma\;=\;({\cal L}_\Gamma, {\cal M}_\Gamma,{\cal B}_\Gamma, E_\Gamma^{(\Phi)}, E_\Gamma^{(U)}, \Phi_\Gamma, \Psi_\Gamma).
\ee
Here ${\cal L}_\Gamma, {\cal M}_\Gamma$ and ${\cal B}_\Gamma$ are three mutually disjoint sets of internal lines of $\Gamma$,
beams of $\Gamma$, and vertices of $\Gamma$, respectively. $E_\Gamma^{(\Phi)}, E_\Gamma^{(U)}$ are maps. 
$E_\Gamma^{(\Phi)}$ assigns the number of external $\Phi$-lines to 
every vertex $v\in {\cal B}_\Gamma$, $E_\Gamma^{(U)}$ assigns
the number of external $U$-lines to every beam. $\Phi_\Gamma$ and $ \Psi_\Gamma$ are incidence relations that assign internal
lines to their endpoint vertices and beams to their internal lines, 
respectively. 
In our former applications of DLCE, lines were 
treated as undirected. So we do here for simplicity, because the 
generalization is easily done. On the other hand it
will be necessary, if later our graphs are considered as so-called 
motifs, which do make use of directed lines, cf. section 3.
We consider $\overline{{\cal B}_\Gamma\times{\cal B}_\Gamma}$ as the set of unordered pairs of vertices $(v,w)$ with
$v,w\in {\cal B}_\Gamma$. Then we have $\Phi_\Gamma: {\cal L}\longrightarrow 
\overline{{\cal B}_\Gamma\times{\cal B}_\Gamma}$. 
We say  $v$ and $w$ are the endpoint vertices of $l\in {\cal L}_\Gamma$ 
if $\Phi_\Gamma(l)=(v,w)$. Similarly
$\Psi_\Gamma: {\cal L}\longrightarrow {\cal M}_\Gamma, l\longmapsto 
\Psi_\Gamma(l)$. 
$\Psi_\Gamma$ associates with each internal line the beam it belongs to. 
Vice versa, a beam connects $m\geq 1$ internal lines, in particular the case
of $m=1$ is allowed where the notion of beam-connectivity becomes redundant. 
As stated above, a line with only one vertex attached is 
an external $\phi$-line, a line with no vertices attached 
is an external $U$-line.
A so-called $\nu$-line is shown in Fig.1.
\begin{figure}[h]
\begin{center}
\setlength{\unitlength}{0.8cm}
\begin{picture}(5.0,3.0)

\epsfig{bbllx=-75,bblly=10,
        bburx=070,bbury=859,
        file=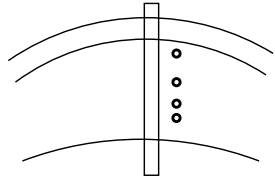,
        scale=0.6}
\end{picture}
\end{center}
\caption{A $\nu$-line made of $\nu$ single lines connected via one beam.}
\end{figure}
\noindent
In a concrete realization the incidence relations $\Phi_\Gamma$ 
and $\Psi_\Gamma$ may be realized as a tuple of matrices
$(\Phi_\Gamma(i,j), \Psi_\Gamma(k,l))$, $i,j\in\{1...n\}$, $k,l\in\{1...m\}$, 
defined in the following way. 
Given a graph $\Gamma$ with $n$ vertices, $m$ internal
lines, $L_\phi$ external $\phi$-lines, $L_U$ external $U$-lines, $r$ beams and a labelling of vertices and internal lines.
$\Phi_\Gamma(i,j)$ is a symmetric $n\times n$-matrix with $\Phi_\Gamma(i,j)$ equal to the number of internal lines 
connecting $i$ and $j$ for $i\not= j$, $i,j \in 1...n$. $\Phi_\Gamma(i,i)$ equals to the number of external $\phi$-lines
attached to vertex $i$. $\Psi_\Gamma$ is a symmetric $m\times m$-matrix 
with $\Psi_\Gamma(k,l)=1$ if internal lines
$k$ and $l$, $k\not= l$, belong to the same beam, $\Psi_\Gamma(k,l)=0$ else, 
$\Psi_\Gamma(k,k)$ equals to the number of 
external $U$-lines attached to line $k$. Notice that two vertices 
may be connected via more than one line, but two 
lines may be connected by at most one beam. (This asymmetry is a remnant 
of the dynamics for which DLCE have been originally proposed.) 
This tuple of matrices, 
representing the incidence relations, would be suited for computer 
implementations of DLCE-graphs. It allows for a computer aided 
algorithmic generation of graphs.

Now we can formulate in a purely algebraic way when two DLCE-graphs 
are topologically equivalent. Two DLCE-graphs
\be
\Gamma_i\;=\;({\cal L}_i, {\cal M}_i, {\cal B}_i, E_i^{(\Phi)}, E_I^{(U)}, \Phi_i, \Psi_i)\qquad \;i=1,2
\ee
are called topologically equivalent if there are three invertible maps
\begin{eqnarray}
&& \phi_{{\cal B}}: {\cal B}_1\longrightarrow {\cal B}_2 \nonumber \\
&& \phi_{{\cal L}}: {\cal L}_1\longrightarrow {\cal L}_2 \nonumber \\
&& \phi_{{\cal M}}: {\cal M}_1\longrightarrow {\cal M}_2 \label{eq.4} 
\end{eqnarray}
between the set of vertices, internal lines, and beams of these graphs $\Gamma_1$ and $\Gamma_2$ such that
\begin{eqnarray}
&& \Phi_2\circ \phi_{{\cal L}}\;=\;\overline{\phi}_{{\cal B}}\circ \Phi_1 \label{eq.a}\\
&& \Psi_2\circ \phi_{{\cal L}}\;=\;\phi_{{\cal M}}\circ \Psi_1 \label{eq.b}
\end{eqnarray}   
and
\begin{eqnarray}
&& E_2^{(\phi)}\circ \phi_{{\cal B}}\;=\;E_1^{(\phi)} \label{eq.c}\\
&& E_2^{(U)}\circ \phi_{{\cal M}}\;=\;E_1^{(U)}\label{eq.d}\;. 
\end{eqnarray}   
Here $\circ $ means decomposition of maps and
\begin{eqnarray}\label{eq.9}
&& \overline{\phi}_{{\cal B}}:\overline{{\cal B}_1\times{\cal B}_1}
\longrightarrow \overline{{\cal B}_2\times{\cal B}_2} 
\nonumber\\
&& \overline{\phi}_{{\cal B}}(v,w)\;\longmapsto (\phi_{{\cal B}}(v), \phi_{{\cal B}}(w)). 
\end{eqnarray}   
For example, (\ref{eq.a}) means that the following composition of maps are equivalent: first assign via $\Phi_1$ 
the endpoint vertices to a given internal line $l_1$ of the first graph $\Gamma_1$ and map them to the corresponding
vertices in $\Gamma_2$ via $\overline{\phi}_{{\cal B}}$, or, 
alternatively, first map the given internal line of the 
first graph $\Gamma_1$ to the corresponding internal line $l_2$ 
of the second graph via $\phi_{{\cal L}}$, and then
associate the endpoint vertices with this line there via $\Phi_2$. 
Similarly, in a shorthand notation, (\ref{eq.b})
refers to the maps $\mbox{line}_{\Gamma_1}\longrightarrow  
\mbox{beam}_{\Gamma_1}\longrightarrow \mbox{beam}_{\Gamma_2}$ or
$\mbox{line}_{\Gamma_1}\longrightarrow  \mbox{line}_{\Gamma_2}
\longrightarrow \mbox{beam}_{\Gamma_2}$. Both orders are equivalent if the
graphs are topologically equivalent. Eq.s (\ref{eq.c}), (\ref{eq.d}) 
state the equivalence of either assigning the external lines 
to a vertex of the first graph, or to the corresponding vertex of the 
second graph.
Fig.2 shows eight topologically inequivalent graphs on the r.h.s. of the map, 
while two topologically equivalent graphs would result 
{} from the graph on the l.h.s. by attaching the beam either to the lower
or to the upper two internal lines. 
(Only the latter graph is shown in the figure.)
The eight graphs on the r.h.s. result from so-called 
admissible vertex resolutions of the graph located most left.
Admissible vertex resolutions will be defined below.

\begin{figure}[h]

\begin{center}
\setlength{\unitlength}{0.8cm}

\begin{picture}(15.0,5.5)

\epsfig{bbllx=-333,bblly=197,
        bburx=947,bbury=738,
        file=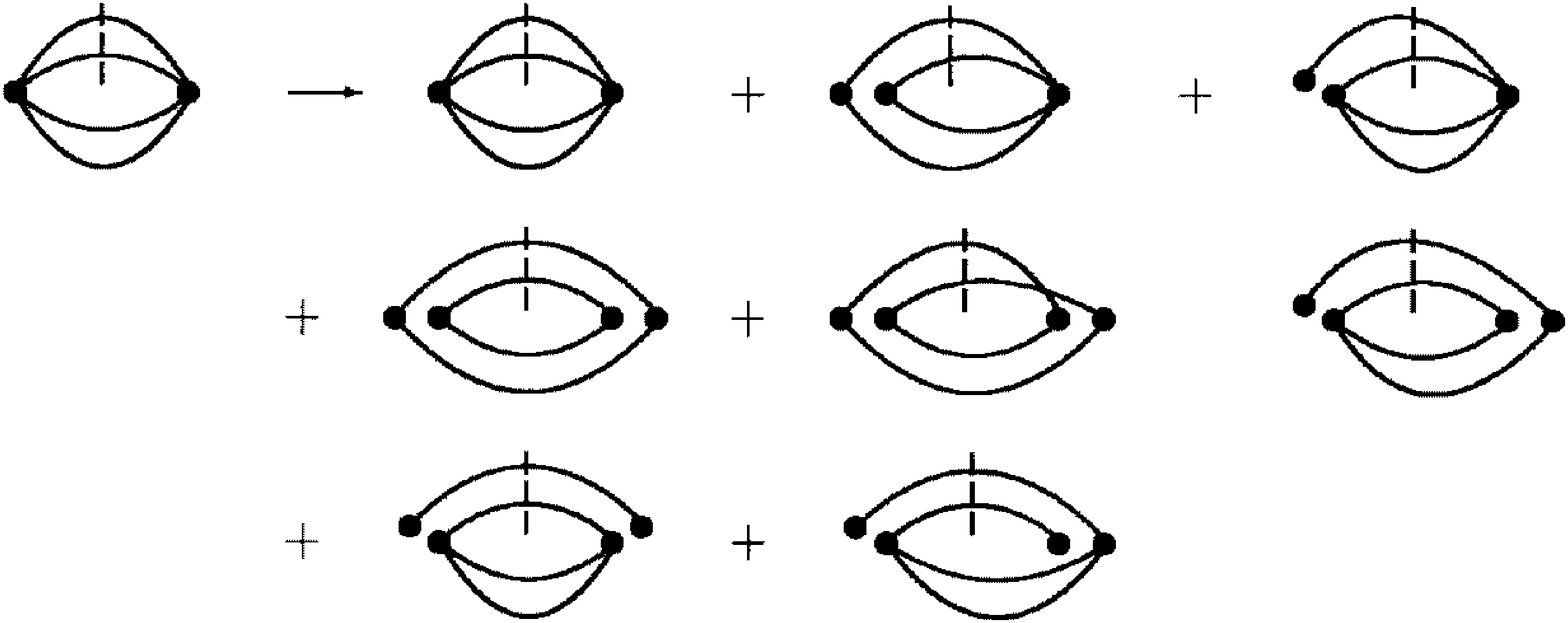,
        scale=0.25}

\end{picture}

\end{center}

\caption{Example of eight admissible, topologically inequivalent vertex resolutions with four internal lines of one common underlying graph. The beams are indicated via dashed lines.}

\end{figure}
\vskip5pt
\noindent
The definition of connectedness of a DLCE-graph with two types of connectivity can be traced back to the notion of 
connectedness of an LCE-graph with only one type of connectivity, for details we refer to \cite{hmoreisz}.
 \vskip5pt
\paragraph{Operations on DLCE-graphs}Apart from operations like adding or removing vertices, lines and beams, with or without the attached structures, 
two operations are of interest in this context, the resolution of vertices and the resolution of beams. Let $\Gamma$
be a DLCE-graph, $v\in {\cal B}_\Gamma$ a vertex with $n$ lines ending upon it, let $\Pi\in {\cal P}({\cal L}_v)$ be any
partition of the set of lines ${\cal L}_v$ ending on $v$. We remove the 
vertex $v$ and draw for every $ P\in \Pi$
a new vertex $v(P)$ so that all lines $l\in P$ enter the vertex $v(P)$ rather than $v$ before its removal. This procedure
is called a vertex resolution of $v$. For an example see Fig.3. Note that this
resolution procedure amounts to a rewiring of lines. 
\begin{figure}[h]

\begin{center}
\setlength{\unitlength}{0.8cm}

\begin{picture}(15.0,7.0)

\epsfig{bbllx=-341,bblly=76,
        bburx=667,bbury=859,
        file=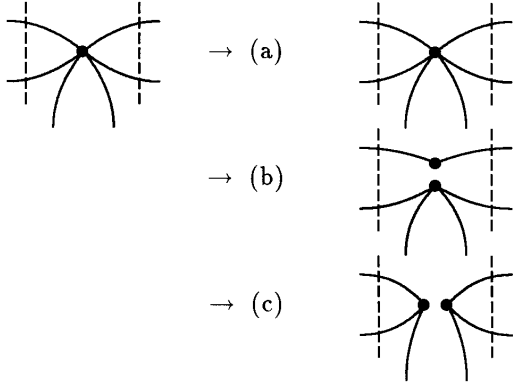,
        scale=0.2}
\end{picture}
\end{center}

\caption{Example of admissible ((a) and (b)) and
non-admissible (c) vertex resolutions.
In particular, the case of the trivial resolution (a)
is admissible. The beams are drawn as dashed lines.}
\end{figure}
It then depends on the dynamical constraints 
whether the resulting graph $\Gamma$
is allowed or not. For example, the graph may become disconnected and fragmentize into several pieces as a result of the
resolution procedure. Such a resolution is forbidden if the considered graphs must be connected. More
generally, a vertex resolution is called admissible if it satisfies all constraints from the dynamics or from the choice
of observables (if the observable is the free energy density, the contributing graphs must be connected because of the
logarithm of the partition function.) 

The same definition applies to the resolution of a beam. The set of all resolution
patterns of a graph $\Gamma$ is obtained by combining any admissible resolution of vertices $v\in {\cal B}_\Gamma$
and any admissible resolution of beams $b\in{\cal M}_\Gamma$ in all possible ways such that the resulting graph satisfies
all characteristic features it should.
\vskip10pt
\noindent
The reason why we choose DLCE-graphs rather than Feynman graphs 
(mentioned in \cite{kri} in a different context of networks or LCE-graphs 
\cite{reisz}
is the following. The notion of resolution of vertices appears quite naturally in DLCE-graphs because graphs may stay 
connected in spite of "resolved" vertices due to the new feature 
of line-connectivity. We remark, however, that also 
in LCE-graphs one meets a similar resolution of vertices 
(although this notion is not introduced there) 
when calculating internal symmetry factors. Let us assume an underlying $O(N)$-symmetry of the system so that one of 
$n$ "colors" (``flavors'', features) may propagate along each line. 
In calculating the internal symmetry factor one looks for all possible 
paths along which color 1, say, can propagate from the input channel
through the graph to yield color 1 on the output 
channel, while a closed loop may carry any of the $N$ colors, 
and only one color can propagate along a line at the
same time. As an example consider the LCE-graph of Fig.4. 
\begin{figure}[h]
\begin{center}
\setlength{\unitlength}{0.8cm}
\begin{picture}(5.0,3.0)

\epsfig{bbllx=250,bblly=10,
        bburx=-670,bbury=-859,
        file=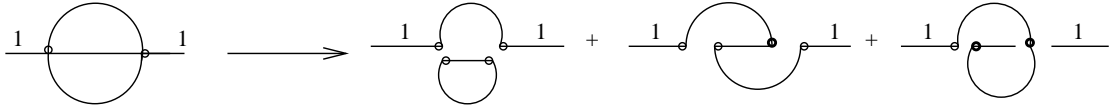,
        scale=0.6}
\end{picture}
\end{center}
\caption{$3\times (N+2)$ possibilities for color 1 to 
propagate through the graph. For further explanations see the text.}
\end{figure}
Color 1 can propagate along the upper line, say $l_1$, leaving $N$ colors for the loop of the remaining lines, say $l_2$ 
and $l_3$, or along $l_1,l_2,l_3$, or
$l_1,l_3,l_2$, or it could choose the intermediate or the lower line first,
yielding $3\times (N+2)$ possibilities all together. 

A resolution of vertices is furthermore encountered if the vertices correspond to knots, and we want to determine all trails or 
states of a given knot \cite{kauffman}. At each crossing of the knot a propagating "bit of information" would have to 
decide where to continue after the crossing.

Finally, in mesoscopic or macroscopic networks one may associate 
with the so-called resolution of vertices rules on ``traffic''
regulation in generic networks of transportation where these rules 
regulate the traffic at crossing points. The rules should
be quite distinct depending on whether energy or information flow is concerned.
 
\section {Proposal of the Complexity Measure}
 
After introducing DLCE-graphs in the last section we are now prepared to 
propose a measure for functional complexity of 
networks. It is defined as 
\begin{equation}
FCM\;:=\;\sum_{i\in(1...N)}^{\prime} PA(\Gamma)\;,
\end{equation}
that is, as
the total number of topologically inequivalent admissible 
resolution patterns $PA$ 
of the DLCE-graph $\Gamma$ of 
that network. The prime stands for the restriction to 
topologically inequivalent and admissible 
patterns. A resolution pattern is 
obtained by allowing for any 
$r_v$ $(0\leq r_v\leq n_v)$ resolutions of vertices and any $r_b$
$(0\leq r_b\leq n_b)$ resolutions of beams, $n_v, n_b$ denoting the total number of vertices and beams, respectively. It is admissible if it is 
compatible with the constraints imposed by the dynamics.
Two resolution patterns are topologically equivalent if there exist the three
invertible maps (\ref{eq.4}) between their associated graphs
$\Gamma_1$ and $\Gamma_2$ which satisfy (\ref{eq.a})- (\ref{eq.9}).

Examples for dynamical constraints are the following:  
\begin{itemize}
\item 
After the resolution of vertices the resulting graph should stay connected.
\item 
Vertices and/or beams have an even number of lines attached.
\item
There are no lines with coinciding endpoints (i.e. no self-lines or tadpoles).
\end{itemize}
\noindent
Some further comments to our proposal are in order. In our definition of the 
complexity measure we count all admissible resolution 
patterns of graphs without assigning
probabilistic weights. In the background of spin glasses the weight was 
given by an analytic
expression proportional to some power $n$ of an expansion parameter $\kappa$ where $n$ equals the number of internal lines in the graph. 
In general the weights for resolution patterns will depend on the imposed
dynamics.

Our conjecture is that the restriction to topologically inequivalent 
resolution patterns projects on the inequivalent
functionalities. Let us consider again the example of the 
graph of Fig.4 
with one input and one output channel
and three topologically inequivalent resolutions. In Fig.4 a) a bit of 
information may be transported via the upper line through
the network while another bit may be stored in the closed loop, in b) and
c) one bit propagates through the graph in a different order, 
without a decoupled loop as in a).

Another example is provided by two motifs observed in foodwebs (Fig.5 a)) 
and in networks of gene regulation (Fig.5  b)).
\begin{figure}[h]

\begin{center}
\setlength{\unitlength}{0.8cm}

\begin{picture}(6.5,4.0)

\epsfig{bbllx=50,bblly=35,
        bburx=-970,bbury=-859,
        file=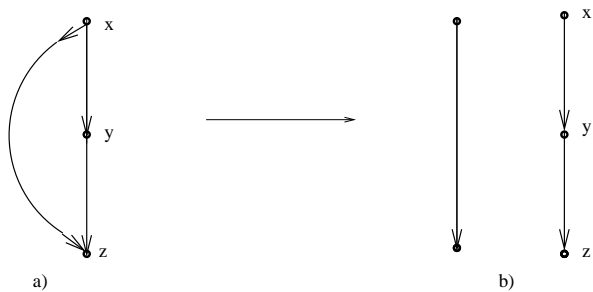,
        scale=0.5}

\end{picture}

\end{center}

\caption{After resolution of its vertices the motif of a feedforward-loop (a) leads to the motif of a three-chain and a two-chain (b).}
\end{figure}
\noindent
According to our definition the directed three-chain  from $x$ to $z$
with a disconnected (here redundant) string (Fig.5 b))
results as a resolution of a feedforward loop (Fig.5 a)) occurring in networks 
of gene regulation, neurons 
and electronic circuits. In this example the interpretation of different 
motifs as resolution patterns of a common
underlying graph appears superficial, because it is more the specific 
difference rather than a common underlying structure
one is interested in. In case of more intricate motifs as in genetic or 
artificial networks the notion of inequivalent
admissible resolution patterns may be a sensible measure for the variety of 
functionalities of these networks.

\section{Background of DLCE}
DLCE-graphs have been proposed in a graphical representation of a systematic generalized high temperature expansion
of $\ln Z$, where $Z$ is the partition function of a certain class of models characterized by an action (Hamiltonian)
of the form
\be\label{eq.10}
S(\phi,U,v) \; = \; \sum_{x \in \Lambda_0}S^0(\phi(x))
             +\sum_{l \in \overline\Lambda_1}S^1(U(l))
             - \frac{1}{2} \, \sum_{x,y \in \Lambda_0} 
               v(x,y)\phi(x)U(x,y)\phi(y) ,
\ee
with couplings
\bea
 &&  v(x,y) \; = \; v(y,x) \; \not= \; 0 
 \qquad   \mbox{for $(x,y)\in\overline\Lambda_1$}, \nonumber \\
 && \mbox{in particular} \; v(x,x) \; = \; 0, \nonumber \\
 && \overline{\Lambda}_1\;=\;\{l=(x,y)\vert v(x,y)\not =0\}.
\eea
\noindent
The field $\phi(x)$ is associated with the sites $x\in\Lambda_0$
where $\Lambda_0$ is the set of all lattice sites, the field $U(l)$
lives on the links (in this paper called lines) $l\in\overline\Lambda_1$,
we write $U(x,y)=U(l)$ for $l=(x,y)$.
In applications to spin glasses 
the $\phi\in \pm1$ are the (fast)
Ising spins and the $U$ $\in {\bf R}$ are the (slow) interactions.
The action is split into two ultralocal parts, $S^0$ depending
on fields on single sites, and $S^1$ depending on fields
on single links $l$.
For simplicity we choose $S^1$ as the same function for all
links $l\in\overline\Lambda_1$. 
The support of $v(x,y)$, here equal to $\overline{\Lambda_1}$, 
need not be restricted to nearest neighbors,
also the precise form of $S^0$ and $S^1$
does not matter
for the generic description of DLCE. $S^0$ and $S^1$ can be any
polynomials in $\phi$ and $U$, respectively.
The only restriction in that context is
the existence of the partition function.

The interaction term $v(x,y)~\phi(x)~U(x,y)~\phi(y)$ contains a
point-link-point interaction and generalizes the 2-point interactions 
$v(x,y)~\phi(x)~\phi(y)$ of usual hopping parameter expansions.
The effective coupling of the $\phi$ fields has its own dynamics
governed by $S^1(U)$, the reason why we have called the new
expansion scheme {\it dynamical} LCE.
Dynamical linked cluster expansions are induced from a Taylor expansion of
$W(H,I,v)=\ln Z(H,I,v)$ about $v=0$, the limit of a completely decoupled
system.
A prominent model which is contained as a special case is the Sherrington-
Kirkpatrick model \cite{sherrington} with spin variables $\phi\in \{-1,+1\}$,
real-valued $U$ and infinite range of connectivity.
Because of the general feature that the couplings $U$ of the "spin" degrees of freedom are not frozen to a specific pattern but have their own dynamics, 
we had to introduce the additional notion of multiple-line 
connectivity, so that 
additional graphs become connected via beams, connecting the internal lines,
and contribute to $\ln Z$.
$\ln Z$ is the generating functional of all DLCE-graphs in the vacuum. $N$-point correlation functions can be derived 
from $\ln Z$ in the usual way, leading to the distinction of internal and external lines, external $\phi$-or $U$-lines,
depending on the choice of $\phi$-or $U$-correlation functions as observables. The dynamical constraints restricting the admissible
vertex resolutions which are specific for the class of models (\ref{eq.10}) 
have been mentioned in the previous section.
The weight of a DLCE-graph is an analytic expression contributing as one term in the expansion of $\ln Z$. Notions like the resolution of vertices were introduced in \cite{hmoreisz} in order to systematically
generate all DLCE-graphs from the more simple LCE-graphs. In spite of resolutions, graphs may stay connected in this
framework if the lines are connected via beams. The assignment 
of DLCE-graphs to generic networks in the present
context is obviously quite different from the original one, but due to their
generic structure not restrictive at all.

\section{Summary and Outlook}
We have proposed a static measure which is sensitive to the topological 
diversity of a graph if we allow for splitting of vertices and induced
rewiring of lines.
To make this measure well defined we used the concept of DLCE-graphs because the 
resolution of vertices was a natural operation in the original framework where DLCE-graphs were used. Moreover DLCE-graphs 
cover more familiar graphs with only line-connectivity as special cases. The actual probability for a certain resolution pattern
of a DLCE-graph depends on the imposed dynamics. In biological applications we expect this measure to be a good indicator
for the functional diversity and therefore the flexibility of a network. 
The subgraph with the highest degree of complexity as defined 
here may be found on top of a hierarchical
system. The various resolution patterns then amount to different functions the network is able to fulfill. Vice versa, interpreting
different patterns, in particular motifs, as result of a common graphical substructure may serve as a classification
of networks with respect to their various functionalities. 
Imposing concrete dynamics on networks with DLCE-graphs will allow to 
relate our complexity measure to others. For a given architecture of the DLCE-graph it then makes sense to ask how many distinct and independent
specific tasks 
the network can perform at the same time. For computational tasks the 
functional complexity may then 
reflect the computational complexity.

%
%


\begin{thebibliography}{9}
\bibitem{wackerbauer} R.Wackerbauer, A.Witt, H.Atmanspacher, J.Kurths, 
and H.Scheingraber, Chaos, Solitons and Fractals {\bf 4}, 133 (1994)
\bibitem{grassberger}P.Grassberger, Int.J.Teor.Phys. {\bf 25}, 907 (1986)
\bibitem{huberman} H.A.Ceccatto, and B.A.Huberman, Physica Scripta {\bf 37}, 145 (1988)
\bibitem{motifs}R.Milo, S.Shen-Orr, S.Itzkovitz, N.Kashtan, D.Chklovskii, and U.Alon, Science {\bf 298}, 824 (2002)
\bibitem{hmoreisz} H.Meyer-Ortmanns and T. Reisz, Int. J. Mod. Phys. {\bf A14}, 947-985 (1999);
                 H.Meyer-Ortmanns and T. Reisz, Eur. Phys. J. {\bf B27}, 549-558 (2002) 
\bibitem{kri} Z.Burda, J.D.Correia, and A.Krzywicki, cond-mat/0104155
\bibitem{reisz} T.~Reisz, Nucl.~Phys. {\bf B450} 569 (1995); 
T.~Reisz, Phys.~Lett. {\bf 360B} 77 (1995).
\bibitem{kauffman}L.H.Kauffman, On Knots, Annals of Mathematics Studies, Princeton University Press (Princeton, New Jersey) 1987, p.132ff
\bibitem{sherrington} D.Sherrington, and S.Kirkpatrick, Phys.Rev.Lett.{\bf 35}, 1972 (1975)

\end{thebibliography}
\end{document}